\documentstyle[12pt,epsf]{article}
\renewcommand{\bar}[1]{\overline{#1}}

\begin{document}

\begin{flushright}
CPT-2000/P.4004\\
USM-TH-92
\end{flushright}

\bigskip\bigskip
\centerline{\large \bf The Quark-Antiquark Asymmetry
of the Nucleon Sea}

\vspace{12pt}

\centerline{\large \bf from $\Lambda$ and $\bar{\Lambda}$ Fragmentation}

\vspace{18pt}

\centerline{\bf Bo-Qiang Ma\footnote{e-mail: mabq@phy.pku.edu.cn}$^{a}$, Ivan Schmidt\footnote{e-mail: ischmidt@fis.utfsm.cl}
$^{b}$, Jacques Soffer\footnote{e-mail: Jacques.Soffer@cpt.univ-mrs.fr}$^{c}$,
Jian-Jun Yang\footnote{e-mail: jjyang@fis.utfsm.cl}$^{b,d}$}

\vspace{8pt}

{\centerline {$^{a}$Department of Physics, Peking University,
Beijing 100871, China,\footnote{Mailing address}}}

{\centerline {CCAST (World Laboratory),
P.O.~Box 8730, Beijing 100080, China,}}

{\centerline {and Institute of Theoretical Physics, Academia Sinica,
Beijing 100080, China}



{\centerline {$^{b}$Departamento de F\'\i sica, Universidad
T\'ecnica Federico Santa Mar\'\i a,}}

{\centerline {Casilla 110-V, 
Valpara\'\i so, Chile}


{\centerline {$^{c}$Centre de Physique Th$\acute{\rm{e}}$orique,
CNRS, Luminy Case 907, F-13288 Marseille Cedex 9, France}}


{\centerline {$^{d}$Department of Physics, Nanjing Normal
University,}}

{\centerline {Nanjing 210097, China}}



\vspace{10pt}
\begin{center} {\large \bf Abstract}

\end{center}

We present a general analysis of the spin transfer
for $\Lambda$ and $\bar{\Lambda}$ production in 
deep-inelastic scattering of polarized
charged leptons on the nucleon, and  
find that
the pattern of different behaviors
of $\Lambda$ and $\bar{\Lambda}$ production
observed by the E665 Collaboration
suggests the possibility of quark-antiquark asymmetries either
in the quark to $\Lambda$ fragmentation functions and/or in the
quark and antiquark distributions of the target proton.
We also point out that the strange-antistrange
asymmetry of the nucleon sea may produce an observable contribution
to the different behaviors 
of $\Lambda$ and $\bar{\Lambda}$ production. 
We find that a softer $\bar{s}(x)$ than $s(x)$
as predicted by the light-cone baryon-meson fluctuation model
of intrinsic quark-antiquark pairs of the nucleon sea might lead
to a reasonable picture.
However, the magnitude is still too small
to explain the E665 data and the conclusion has also strong model-dependence.
This may suggest the importance of quark-antiquark asymmetry 
in the quark to $\Lambda$ fragmentation functions, provided that the E665 
data are confirmed.

\vfill

\centerline{PACS numbers: 14.20.Jn, 12.38.Bx, 13.87.Fh, 13.88.+e}

\vfill
\newpage

It is well known that the production of $\Lambda$ and 
$\bar{\Lambda}$  in deep-inelastic scattering (DIS) 
of lepton on the nucleon may provide
information on the quark content of the target nucleon 
\cite{Lu95,Ell96}, as well
as on the quark to $\Lambda$ fragmentation functions 
\cite{Jaf96,Ma99}. The idea that the fragmentation of the 
$\Lambda$ hyperon in DIS of a charged lepton on a nucleon
target can supply information concerning the strange content
of the nucleon was originally proposed in Refs.~\cite{Lu95,Ell96}.
There are four different combinations
of the polarizations of the charged lepton beam and the
nucleon target: \\
i) both the lepton beam and the nucleon target
are unpolarized; \\
ii) the nucleon target is polarized while the lepton beam
is unpolarized \cite{Lu95,Ell96,Jaf96}; \\
iii) both the lepton beam and the nucleon target
are polarized \cite{Lu96}; \\
iv) the lepton beam is polarized while the
nucleon target is unpolarized \cite{Jaf96}.\\
These different combinations provide different information 
concerning the 
quark distributions and quark to $\Lambda$ fragmentation functions. 
It is suggested in Ref.~\cite{Jaf96} that there are still
large uncertainties in the quark to $\Lambda$ fragmentation
function, and it is practically more urgent to measure the 
$\Lambda$ fragmentation functions before using the $\Lambda$ 
fragmentation to probe the quark content of the nucleon. 
Indeed, some symplifying assumptions about the quark to $\Lambda$
fragmentation functions were found to be of little
predictive power when applied to $\Lambda$ production
in $e^+e^-$ annihilation process at the $Z$ resonance 
\cite{Bor98,Flo98b}, and to semi-inclusive $\Lambda$ production 
of polarized charged lepton DIS process on the nucleon target 
\cite{Kot98}.

However, there have been recent progress \cite{MSY2,MSY3,MSSY5}
in order to understand the
quark to $\Lambda$ fragmentation functions by connecting
them with the quark distributions inside the $\Lambda$
by the Gribov-Lipatov relation (GLR) \cite{GLR}:
\begin{equation}
D_q^h(z) \sim q_h(x)~,
\label{GLR}
\end{equation}
where $D_q^h(z)$ is the fragmentation function
for a quark $q$ splitting into a hadron $h$ with
longitudinal momentum fraction $z$, and $q_h(x)$
is the quark
distribution for finding the quark $q$ inside the hadron
$h$ carrying a momentum fraction $x$.
$D^h_q$ and $q_h$ depend also on the energy scale $Q^2$, and this
relation holds, in principle, in a certain $Q^2$ range and in
leading order approximation. It is shown recently \cite{Blu00} that
the Gribov-Lipatov relation
is also
verified to hold in leading order for the space- and time-like splitting
functions of QCD.  
Moreover, although Eq.~(\ref{GLR}) is only
valid at $x \to 1$ and $z \to 1$, it provides a reasonable
guidance for a phenomenological parametrization of the various
quark to $\Lambda$ fragmentation functions.
We are encouraged to find that the predictions of the quark
to $\Lambda$ fragmentation functions in
an SU(6) quark-diquark model \cite{Ma96} and in a pQCD based 
model \cite{Bro95} are in good agreement with
the experimental data on $\Lambda$ production in both the
$e^+e^-$ annihilation process at the $Z$ resonance \cite{MSY3}
and in polarized positron beam DIS on a nucleon target 
\cite{MSY2,MSSY5}. Thus we have at least some reasonable 
parametrizations of quark to $\Lambda$ fragmentation
functions, though there are still large uncertainties in the flavor
and spin decompositions of these fragmentation functions.

For a longitudinally polarized charged
lepton beam and an unpolarized nucleon target, 
the longitudinal spin transfer to the $\Lambda$
is given in the quark parton model by
\cite{Jaf96}
\begin{equation}
A^{\Lambda}(x,z)= \frac{\sum\limits_{q} e_q^2 [q^N(x,Q^2) \Delta
D_q^\Lambda(z,Q^2) + ( q \rightarrow \bar q)]}{\sum\limits_{q} e_q^2 [q^N (x,Q^2)
D^\Lambda_q(z,Q^2) + ( q \rightarrow \bar q)]}~.
\label{DL}
\end{equation} 
Here $y=\nu/E$, $x=Q^2/{2M_N \nu}$, and $z=E_\Lambda /\nu$, where
$q^2=-Q^2$ is the squared four-momentum transfer of the
virtual photon, $M_N$ is the proton mass, and $\nu$, $E$, and $E_{\Lambda}$
are the energies of the virtual photon, the target nucleon,
and the produced $\Lambda$ respectively, in the target rest 
frame; $q^N(x,Q^2)$ is the quark distribution for the
quark $q$ in the nucleon, $D_q^\Lambda (z,Q^2)$ is the
fragmentation function for $\Lambda$ production from quark  $q$,
$\Delta D _q^\Lambda (z, Q^2) $ is the corresponding longitudinal
spin-dependent fragmentation function, and $e_q$ is the quark
charge in units of the elementary charge $e$. 
In a region where $x$
is large enough, e.g. $x > 0.2$,  one can neglect the
antiquark contributions in Eq.~(\ref{DL}), and  probe only the 
valence quarks of the target nucleon. On the contrary, if $x$ is 
much smaller, one is probing the sea quarks and therefore the
antiquarks must be considered as well.
For $\bar\Lambda$ production the spin transfer
$A^{\bar\Lambda}(x,z)$ is obtained from Eq.~(\ref{DL}) by replacing
$\Lambda$ by $\bar\Lambda$. The $\Lambda$ and $\bar\Lambda$
fragmentation functions are related since we can safely assume
matter-antimatter symmetry, {\it i.e.}
$D^\Lambda_{q,\bar{q}}(z)=D^{\bar{\Lambda}}_{\bar{q},q}(z)$ and
similarly for $\Delta D^\Lambda_{q,\bar{q}}(z)$.

Recently, the HERMES Collaboration at DESY reported the
result of the longitudinal spin transfer to the $\Lambda$ in
polarized positron DIS on the proton \cite{HERMES}.
Also the E665 Collaboration
at FNAL measured the $\Lambda$ and $\bar{\Lambda}$ spin transfers
from muon DIS \cite{E665}, and they observed very different
behaviour for $\Lambda$ and $\bar{\Lambda}$ polarizations.
The E665 data for the spin transfer are
presented as function of the Feynman variable $x_F$, although 
$x_F \approx z$ is a good approximation in the kinematic range of
the E665 experiment \cite{MSSY5}.
Strictly speaking, the magnitude of the 
measured spin transfer Eq.(\ref{DL}) should
be less than unity; thus the E665 data, whose 
range of magnitude for the measured spin transfer is
larger than unity, are
of poor precision. But the different behaviors of the $\Lambda$ and
$\bar{\Lambda}$ spin transfer might still be a realistic effect.
Both the HERMES data and the E665 data are measured for
$x_F > 0$, which corresponds to the current
fragmentation region. Thus it is natural to try to 
understand the data from the viewpoint of current fragmentation, 
rather than target fragmentation as suggested by E665.
We will focus our attention on
the different behavior of the $\Lambda$ and
$\bar{\Lambda}$ spin transfer in the E665 data.

It is interesting to notice that the fragmentation functions
from the quark-diquark model \cite{MSY2,MSY3} 
can give very good descriptions
of both the data of $\Lambda$ fragmentations in $e^+e^-$ 
annihilation at the $Z$ resonance \cite{MSY3}, and in polarized 
positron DIS on the unpolarized proton by HERMES \cite{MSY2},
with only naive parameters
without any adjustment. Although the Gribov-Lipatov relation
should be of poor validity at small $x$, 
the fragmentation functions obtained by using it in the 
quark-diquark model seem to give a reasonable relation between 
different quark to $\Lambda$ fragmentation functions. We would 
like to mention that the fragmentation functions derived in a
quark-diquark picture \cite{Nza95} and in an MIT model framework
\cite{Bor99b} arrived at similar 
qualitative results as in Ref.~\cite{MSY2},
although the explicit shapes are not the same.
Therefore we first use the fragmentation functions
from the quark-diquark model as input in order
to calculate the spin transfer,
Eq.~(\ref{DL}), for the $\Lambda$ and $\bar{\Lambda}$
production.

We use the recent CTEQ5 parametrizations as input
for the quark distributions of the nucleon \cite{CTEQ5}. 
In Fig.~1(a)
we present the calculated results 
for the spin transfer
of $\Lambda$ and $\bar{\Lambda}$, and compare the results
with the the HERMES and E665 data. We notice that the
calculations show a trend of increasing positive
polarization with increasing $z$ that seems to be suggested
by the data. This supports the prediction of positive
polarized $u$ and $d$ quarks inside $\Lambda$ at large
$x$ \cite{MSY2,MSY3}. However, the calculations 
cannot produce a difference
of the spin transfers for $\Lambda$ and $\bar{\Lambda}$ as
observed by E665. Our results are
also in qualitative agreement with a Monte Carlo simulation
based on the naive quark model and a model with SU(3)
symmetry \cite{Ash99}, where the different behavior of
spin transfer for $\Lambda$ and $\bar{\Lambda}$ are 
not predicted.  Although the effect due to target
fragmentation has been suggested as a possible
mechanism for the E665 different behavior of $\Lambda$
and $\bar{\Lambda}$ production \cite{E665}, the kinematic
region for $\Lambda$ and $\bar{\Lambda}$ production
corresponds to $x_F > 0$, which is
the current fragmentation region \cite{MSSY5}. Thus we need to 
find a new mechanism for the
different behavior of $\Lambda$ and $\bar{\Lambda}$
production. The possibility of a quark-antiquark asymmetry
in the fragmentation functions has been investigated in
Ref.~\cite{MSSY5}, and the purpose of this paper is to 
investigate possible asymmetries in the quark 
distributions of the nucleon target.

\begin{figure}
\begin{center}
\leavevmode {\epsfysize=7.5cm \epsffile{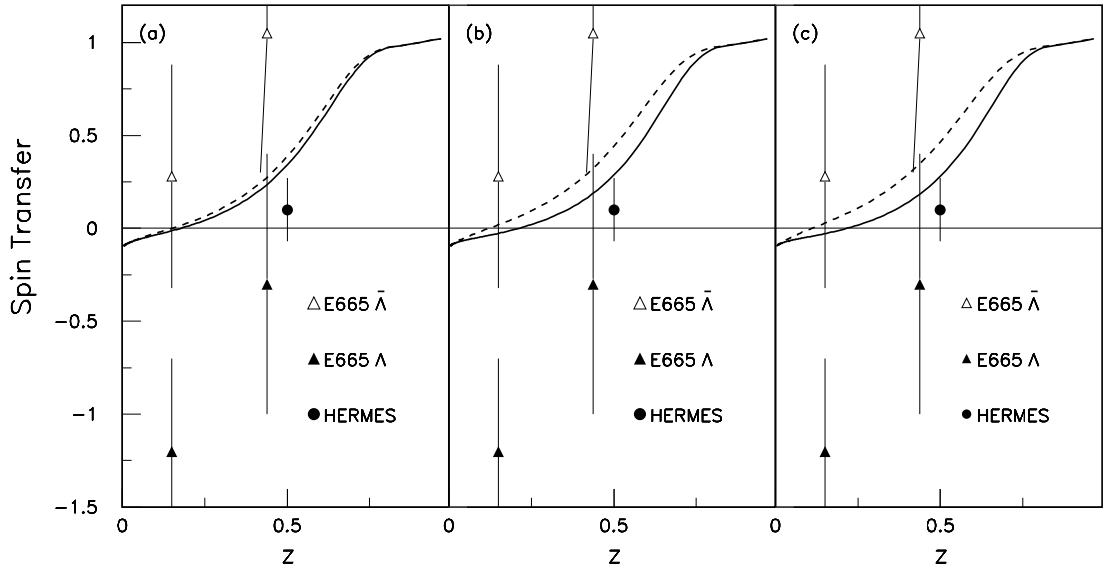}}
\end{center}
\caption[*]{\baselineskip 13pt The $z$-dependence of the $\Lambda$
and $\bar{\Lambda}$  spin transfer in polarized charged lepton DIS
on the nucleon. The solid and dashed curves
correspond to the calculated results of  $\Lambda$
and $\bar{\Lambda}$ spin transfers with: (a) the CTEQ5 set 1
parametrization of quark distributions \cite{CTEQ5}; (b) the modified
CTEQ5 set 1 quark  distributions including only
strange-antistrange asymmetry Eq.~(\ref{cases}); 
(c) the modified CTEQ5 set 1 quark
distributions including an additional contribution of 
$\bar{d}(x) > d(x) $ in the target proton Eq.~(\ref{caseud}).
The quark to $\Lambda$ fragmentation functions are predicted by the
light-cone SU(6) quark-diquark model, and the average value of 
the Bjorken variable is chosen as $x=0.005$, corresponding to the
E665 average value. }\label{mssy6f1}
\end{figure}


Let us consider the spin transfer for $\Lambda$
and $\bar \Lambda$ production in a region where
the sea dominates, namely where the Bjorken $x$
is rather small, like in the E665 experiment, 
which has $\left<x_B\right>$=0.005.
The E665 data indicates that in this region 
and for $z$ between 0.1 and 0.5, one has
\begin{equation}
A^{\bar \Lambda}(x,z) >> A^\Lambda(x,z).
\label{ST}
\end{equation}
Let us consider several possible situations:\\
1) The sea is fully symmetric, namely $q(x)= \bar q(x)$, for all 
flavors $u,d,s$. This implies clearly  $A^{\bar \Lambda}(x,z) = 
A^\Lambda(x,z)$, which contradicts the data.\\
2) The sea is not fully symmetric, and in this case one can 
consider several scenarios: 
\\
(a) One flavor dominates, for example $u$-quark. Let us define
\begin{equation}
\Delta \bar Q = \bar q(x)[\Delta D^{\Lambda}_q (z)+\Delta D^{\Lambda}_{\bar q }(z)]~~,
~~\bar Q = \bar q(x)[D^{\Lambda}_q (z)+ D^{\Lambda}_{\bar q }(z)]~.
\end{equation}
In this case, if the sea is symmetric we are back to case 1) above
and Eq.~(\ref{ST}) cannot be satisfied. If the sea
is not symmetric, namely $u=\bar u + \epsilon$, we have
\begin{equation}
A^{\Lambda}= \frac{ \epsilon \Delta D^{\Lambda}_u + 2\Delta \bar U}{ \epsilon D^{\Lambda}_u + 2\bar U} ~,
\end{equation}
and
\begin{equation}
A^{\bar \Lambda}= \frac{ \epsilon \Delta D^{\Lambda}_{\bar u} + 2\Delta \bar U}{ \epsilon D^{\Lambda}_{\bar u} + 2\bar U} ~.
\end{equation}
Let us try to see what conditions one must have in order to
fulfill Eq.~(\ref{ST}).
It seems clear that $D^{\Lambda}_u >> D^{\Lambda}_{\bar u}$,
therefore if $\epsilon > 0$ (which is the case for $x=0.005$
in present parametrizations of quark distributions), Eq.~(\ref{ST})
will be satisfied provided
\begin{equation}
\Delta D^{\Lambda}_{\bar u} >> \Delta D^{\Lambda}_u ~,
\end{equation}
a condition which is assumed and discussed in \cite{MSSY5} as
a possibility to explain the different behaviors of $\Lambda$ and
$\bar{\Lambda}$ productions in the E665 data.
\\
(b) All three flavors contribute but only one is asymmetric, say 
$u$, whereas $d=\bar d$ and $s=\bar s$. This case is similar
to the previous one, since we have
\begin{equation}
A^{\Lambda}= \frac{ 4\epsilon \Delta D^{\Lambda}_u + X}
{ 4\epsilon D^{\Lambda}_u + Y} ~,
\end{equation}
and
\begin{equation}
A^{\bar \Lambda}= 
\frac{ 4\epsilon \Delta D^{\Lambda}_{\bar u} + X}
{ 4\epsilon D^{\Lambda}_{\bar u} + Y} ~,
\end{equation}
where $X= 8\Delta \bar U + 2\Delta \bar D + 2\Delta \bar S $ 
and $Y= 8\bar U + 2\bar D + 2\bar S $.
We reach the same conclusion as above, and since the asymmetric 
flavor can be either $d$ or $s$, it seems that one should have 
more generally \begin{equation}
\Delta D^{\Lambda}_{\bar q} >> \Delta D^{\Lambda}_q ~,
\end{equation}
a condition which has been discussed and considered in \cite{MSSY5}.
Remember that positivity implies $D^{\Lambda}_{\bar q} \geq \Delta D^{\Lambda}_{\bar q}$, so it means that one should have
a strong bound on $\Delta D^{\Lambda}_q$, namely
\begin{equation}
D^{\Lambda}_{\bar q} >> \Delta D^{\Lambda}_q ~.
\end{equation}

However, the situation will be different
in case we have $\epsilon <0$, which means that $\bar{q}(x) > q(x)$.
The strange quark-antiquark asymmetry predicted by the light-cone
baryon-meson fluctuation model \cite{Bro96} introduces such
a behavior for the strange quarks and antiquarks. Thus the pattern
of the difference in the $\Lambda$ and $\bar{\Lambda}$ productions
in the E665 data could suggest a possibility of $\bar{q}(x) > q(x)$
in the target proton.

The CTEQ parametrizations of quark distributions are based on data
of various structure functions from different DIS processes 
obtained in the last three decades. The light-flavor $u$ and $d$
content of the nucleon is well constrained and the uncertainties
are not big, though there are still a number of phenomenological 
anomalies related to the spin and flavor content of the nucleon 
sea \cite{Bro96}. However, the strange content of the nucleon is
less known than the light-flavor $u$ and $d$ quarks. In the CTEQ
parametrizations, identical strange and antistrange quark
distributions are assumed. 
However, it is pointed out in
Ref.~\cite{Bro96} that within the allowed errors,
the CCFR data of $s(x)/\bar{s}(x)$ \cite{CCFR95} 
does not rule out
a strange-antistrange asymmetry, as suggested by the
light-cone baryon-meson fluctuation model \cite{Bro96}. 
Moreover, this light-cone baryon-meson fluctuation model
of intrinsic quark-antiquark ($q\bar{q}$) pairs in the
nucleon sea suggests a soft $\bar{s}(x)$
compared to $s(x)$
(i.e., $\bar{s}(x) > s(x)$ at small $x$ and {\it vice versa}
at large $x$). Remember that a softer $\bar{s}(x)$ than $s(x)$
was predicted by Burkardt and
Warr \cite{Bur92} from the chiral Gross-Neveu model
at large $N_c$ in the light-cone formalism.
It is also pointed out in Ref.
\cite{Bro96} that the conflict
between two different determinations
of the strange quark distributions \cite{CTEQ93,Ma96a}
could be a
phenomenological support for $s(x) \neq \bar{s}(x)$,
or more explicitly, a softer $\bar{s}(x)$ compared to $s(x)$.
Another phenomenological support for a softer $\bar{s}(x)$
is also suggested by Barone, Pascaud, and Zomer \cite{Bar99}
from a global QCD analysis
of structure functions, including neutrino DIS
data. More recently, Buccella, Pisanti, and Rosa
\cite{Buc00} found, from
their analysis of the new CCFR data on structure functions
at small $x$, an alternative independent support
for a softer $\bar{s}$, in agreement with the prediction of
Ref.~\cite{Bro96}. 
Therefore we can check the
possibility that the different behavior of spin transfer for
$\Lambda$ and $\bar{\Lambda}$ come from the strange-antistrange
asymmetry of the nucleon sea. 
Indeed, the E665 data are measured
corresponding to the quark distributions of the nucleon in the
Bjorken variable range $0.0001 < x < 0.1 $ with $\left< x
\right>=0.005$, where the antiquark distributions are of the same
order as those of the quark distributions. From the light-cone
baryon-meson fluctuation model \cite{Bro96} we know that the
antistrange quark distribution could be as big as more than
two times that of the strange quark distribution at small $x$.
Therefore we can modify the
strange and antistrange quark distributions of the CTEQ
parametrization and check the role played by the
strange-antistrange asymmetry for the $\Lambda$ and
$\bar{\Lambda}$ productions.

We choose the values of the quark distributions with
quark-antiquark asymmetry of the nucleon sea at $x=0.005$ as

\begin{equation}
\begin{array}{clllc}
u=u_0 + \delta u=50.32; 
\\
d=d_0 + \delta d=45.41; 
\\
s=s_0 + \delta s=17.11-8;
\\
\bar{u}=\bar{u}_0 + \delta \bar{u}=33.55; 
\\
\bar{d}=\bar{d}_0 + \delta \bar{d}=35.29; 
\\
\bar{s}=\bar{s}_0 + \delta \bar{s}=17.12+8;
\\
\end{array}
\label{cases}
\end{equation}
where $q_0$ and $\delta q$
($q=u, d, s, \bar{u}, \bar{d}, \bar{s}$)
are the quark distributions of CTEQ parametrization at $x=0.005$
and the corresponding modifications, respectively. In principle 
we can also introduce the nucleon sea quark-antiquark asymmetry in the 
light flavor $u$ and $d$ quarks, but we still 
have no phenomenological evidence for doing this.
In comparison, there are large uncertainties concerning the 
strange and antistrange content of the nucleon sea. Therefore we
first check the role played by the
strange-antistrange asymmetry in the nucleon
target for the difference between $\Lambda$ and $\bar{\Lambda}$
productions in polarized charged lepton scattering on the nucleon.
We present in Fig.~1(b) of the
calculated results with strange-antistrange asymmetry.
It is interesting to
find that the strange-antistrange asymmetry of the nucleon sea
predicted by the light-cone baryon-meson fluctuation model can indeed
produce a trend for the different behaviors of the spin transfers for
$\Lambda$ and $\bar{\Lambda}$ production observed by E665,
though the magnitude is still not enough to explain the data.

 From Eq.~({\ref{DL}), we find that the different strange
and antistrange quark distributions of the nucleon sea
are the reason for the different $\Lambda$ and $\bar{\Lambda}$
production. The $u$ and $d$ ($\bar{u}$ and $\bar{d}$) quarks
mainly contribute to the background of the $\Lambda$
($\bar{\Lambda}$) production. In the quark-diquark model
of the quark to $\Lambda$ fragmentation functions
\cite{MSY2,MSY3}, the quark helicities of $u$ and $d$ quarks
have almost zero net contribution in the whole $x$ range $0 \to 1$.
But this does not seem to be true from the SU(3) symmetry
argument that the $u$ and $d$ quarks may have net
helicilities of the order of -0.2 \cite{Bur93}.
Therefore the absolute
values of the spin transfers at small $z$ might not be
correctly predicted by the quark-diquark model parametrization
of quark to $\Lambda$ fragmentation functions \cite{MSY2,MSY3}.
The interesting aspect is the difference of the $\Lambda$
and $\bar{\Lambda}$ productions from the strange-antistrange
asymmetry of the nucleon sea. From Fig.~1(b) we notice that
the magnitude of the difference can
be the order of $0.25$, which should be large enough to
cause an observed difference in the measurements
of $\Lambda$ and $\bar{\Lambda}$ productions in polarizied
charged lepton DIS process on the nucleon. This shows
that the strange quark content of the nucleon
could be probed after we carefully consider
the effect of $u$ and $d$ quarks and antiquarks of the nucleon,
and of various quark to $\Lambda$ fragmentation functions.

We would like to mention that the above conclusion
depends on the specific forms of the quark
to $\Lambda$ fragmentation functions used as
input for the spin transfer. We also present
in Fig.~2(a) and (b) the calculated results
with and without strange-antistrange
asymmetry of the nucleon sea, but with the fragmentation
functions from a pQCD based model \cite{Bro95} which is also good
in describing the data of $\Lambda$ production in
$e^+e^-$ annihilation at the $Z$ resonance \cite{MSY3} and in the
polarized positron DIS on the proton by HERMES \cite{MSY2,MSSY5}.
We notice that the
difference between the spin transfer for $\Lambda$
and $\bar{\Lambda}$ is small in this situation.
However, the sea quark-antiquark asymmetry in the
quark and antiquark fragmentations to the $\Lambda$
has been found to be an alternative possibility for
the different behaviors of the spin tranfer
for $\Lambda$ and $\bar{\Lambda}$ production \cite{MSSY5}.

\begin{figure}
\begin{center}
\leavevmode {\epsfysize=7.5cm \epsffile{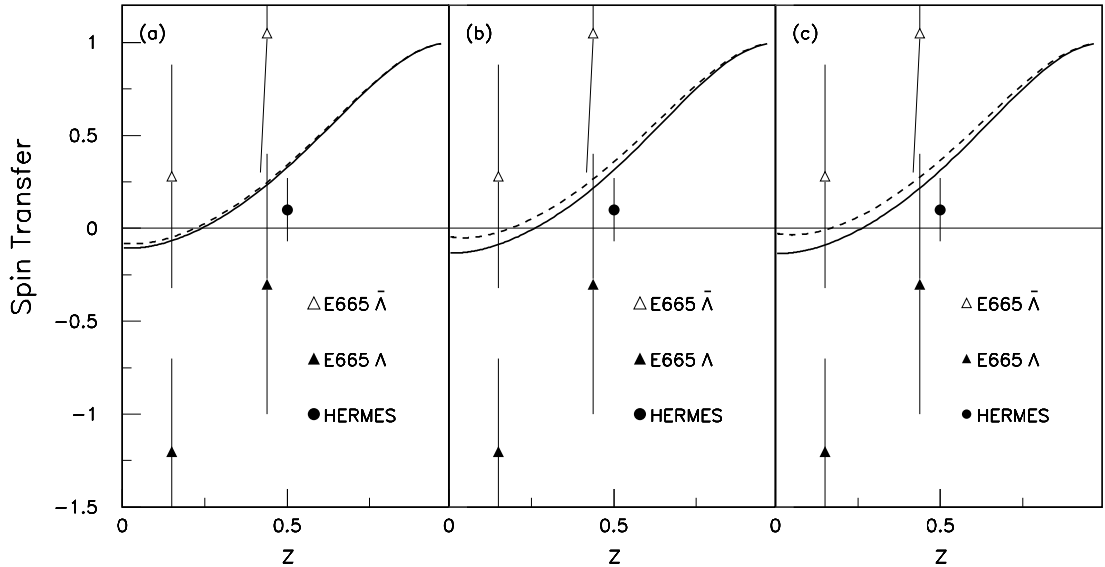}}
\end{center}
\caption[*]{\baselineskip 13pt 
Same as Fig.~\ref{mssy6f1}, but the quark to
$\Lambda$ fragmentation functions are predicted by a 
perturbative QCD (pQCD) based model. }\label{mssy6f2}
\end{figure}

The above discussion helps us to understand why
the strange quark-antiquark asymmetry can provide some contribution
to the different behavior of $\Lambda$ and $\bar{\Lambda}$ productions.
This possibility can only manifest itself in the specific
situation when the strange quarks
and antiquarks are important and $\bar{s}(x) > s(x)$.
Strictly speaking, there have been many experimental data
related to the $u$ and $d$ quark and antiquark distributions
so that there should be less freedom to introduce $\epsilon <0$
for the $u$ and $d$ quarks.  However, we notice that a possibility
of $\bar{d}(x) > d(x)$ is not completely forbidden in the baryon-meson
fluctuation picture to understand the Gottfried sum rule violation
\cite{Bro96}. Therefore we consider another case
with an additional contribution of $\bar{d}(x) > d(x)$ in the target
proton:
\begin{equation}
\begin{array}{clllc}
u=u_0 + \delta u=50.32 +8;
\\
d=d_0 + \delta d=45.41 -8;
\\
s=s_0 + \delta s=17.11-8;
\\
\bar{u}=\bar{u}_0 + \delta \bar{u}=33.55 -8;
\\
\bar{d}=\bar{d}_0 + \delta \bar{d}=35.29 +8;
\\
\bar{s}=\bar{s}_0 + \delta \bar{s}=17.12+8.
\end{array}
\label{caseud}
\end{equation}
The calculated spin transfer for $\Lambda$ and $\bar{\Lambda}$
production are presented in Figs.~\ref{mssy6f1}(c)
and \ref{mssy6f2}(c). We find that $\bar{d}(x) > d(x)$ could only provide
a very small contribution to the different behaviors
of $\Lambda$ and $\bar{\Lambda}$ productions 
with fragmentation functions from both 
the quark-diquark model and the pQCD based
model. This is due to the $u$ quark dominance and it also suggests
that the $\Lambda$ and $\bar{\Lambda}$ production at small $x$
is not sensitive to the $d$ and $\bar{d}$ quark distributions, but 
might be sensitive to the $s$ and $\bar{s}$ quark distributions,   
although there is strong model-dependence in this conclusion.


 From the above general analysis we find 
that the different behaviors of $\Lambda$ and $\bar{\Lambda}$
productions could be due to either the quark-antiquark asymmetries 
in the quark fragmentations and/or in
the nucleon sea. The strange-antistrange asymmetry 
of the nucleon could provide
a contribution to the observed difference of  
$\Lambda$ and $\bar{\Lambda}$
productions, but the magnitude is still too small to explain the
data. It thus forces us to 
consider the importance of the quark-antiquark asymmetry 
in the quark fragmentations if the E665 data are confirmed.

However, due to large uncertainties in the data 
and in the various quark to
$\Lambda$ fragmentation functions, it is still too early for us to
arrive at some definite conclusion other than to suggest
some interesting possibilities for further study.
Thus we still need further efforts in order to reduce the 
uncertainties in the spin and flavor structure of various quark 
to $\Lambda$ fragmentation functions.
We know that the $\Lambda$ and $\bar{\Lambda}$ fragmentation
in neutrino (antineutrino) DIS processes \cite{Ma99,MSSY5}, and
the different combinations of beam and target polarizations 
in the charged lepton DIS on the nucleon,
can provide further insight on this issue,
in addition to the $\Lambda$ ($\bar{\Lambda}$)
fragmentation in the $e^+e^-$ annihilation near the $Z$
resonance \cite{MSY3}. We expect further theoretical and experimental
work to push forward progress in this direction.

{\bf Acknowledgments: } This work is partially supported
by Fondecyt (Chile) postdoctoral fellowship 3990048, by the
cooperation programmes Ecos-Conicyt and CNRS- Conicyt between
France and Chile, by Fondecyt (Chile) grant 1990806 and by a
C\'atedra Presidencial (Chile), and
by National Natural Science
Foundation of China under Grant Numbers 19605006, 19875024,
19775051, and 19975052.

\end{document}